\documentclass{appolb}
\usepackage{graphicx}
\usepackage{amsmath,amssymb}
\usepackage{bm}
\newcommand{\trento}{T\raisebox{-.5ex}{R}ENTo}
\begin{document}
\title{Photon Production in High-Energy Heavy-Ion Collisions: 
Thermal Photons and Radiative Recombination%
\thanks{Presented at the XXIXth International Conference on Ultra-relativistic Nucleus-Nucleus Collisions (QM2022).}%
}
\author{Hirotsugu Fujii$^1$, Kazunori Itakura$^{2,3}$, Katsunori Miyachi$^4$, \\Chiho Nonaka$^{4,5,6}$
\address{$^1$Institute of Physics, University of Tokyo, Komaba, Tokyo 153-8902, JAPAN \\
$^2$Nagasaki Institute of Applied Science, Nagasaki 851-0193, JAPAN\\
$^3$KEK Theory Center, IPNS, KEK, Tsukuba 305-0801, JAPAN \\
$^4$Department of Physics, Nagoya University, Nagoya 464-8602, JAPAN\\
$^5$Physics Program, Graduate School of Advanced Science and Engineering, Hiroshima University, Higashi-Hiroshima 739-8526, JAPAN\\
$^6$Kobayashi-Maskawa Institute for the Origin of Particles and the
Universe (KMI), Nagoya University, Nagoya 464-8602, JAPAN
}
}
\maketitle
\begin{abstract}
We present a comprehensive analysis of photon production at RHIC and the LHC, proposing  
radiative hadronization as an additional photon source in high-energy heavy-ion collisions. 
For thermal photon, we perform relativistic viscous hydrodynamic calculation 
with event-by-event fluctuations. 
\end{abstract}
  
\section{Introduction}
Direct photon is considered as an important probe to extract information of quark-gluon plasma (QGP)
 and hot hadronic matter through space-time evolution of high-energy heavy-ion collisions.  
Measurements of the photons in relativistic heavy-ion collisions have been performed both at RHIC and the LHC, and 
large yields of the transverse momentum spectra and strong elliptic flow of photons are reported. 
Any theoretical model so far seems to be incapable of explaining the photon data adequately. 
The large yield could be attributed to early stage 
of the evolution with higher temperatures, while the strong collective flow prefers large photon emission at later stage when momentum anisotropy of QGP is well developed. 
This situation is called as the “direct photon puzzle”.

In this paper, we propose another source of photon production which has been overlooked and is inherent to late stages of the QGP time evolution: it is photon radiation at hadronization of QGP,  which is, in fact, natural from the viewpoint of ordinary electromagnetic plasmas. 
In addition, we carry out numerical computation of  thermal photons from the relativistic 
viscous hydrodynamic model and prompt photons and show a comprehensive study of direct photons 
at RHIC and the LHC. 

However, we should note that there is a reservation about the experimental results because the large yield of photons measured by PHENIX has not been confirmed by STAR. Also, the deviation between thermal photons 
from a relativistic hydrodynamic model and the experimental data is not so large at the LHC.

\section{Photon production in High-Energy Heavy-Ion Collisions}

\subsection{Thermal Photons from Hydrodynamics}
For computation of thermal photon, we use a relativistic viscous hydrodynamic model which is composed  of 
\trento\  for initial condition, hydrodynamic expansion including shear and bulk 
viscosities and final state interactions described by UrQMD \cite{Okamoto:2017rup}. 
The temperature dependence of shear and bulk viscosities is included. 
All the parameters of the model are tuned  from rapidity distribution, $p_T$ spectra and elliptic flow of charged hadrons \cite{Okamoto:2017rup,Miyachi-Nonaka}.

Thermal photons are emitted from the QGP and the hadronic phases during hydrodynamic 
expansion. 
We combine the contributions from the QGP and the hadronic phase by smooth interpolation formula as
\begin{equation}
  \epsilon \frac{dR_{\rm th}^\gamma}{d^3 k}
  = \frac{1}{2} \left ( 1- \tanh \frac{T-T_c}{\Delta T} \right )
  \epsilon \frac{dR_{\rm had}^\gamma}{d^3 k}
  +
  \frac{1}{2} \left ( 1 + \tanh \frac{T-T_c}{\Delta T} \right )
  \epsilon \frac{dR_{\rm QGP}^\gamma}{d^3 k}
  \label{eq_had_QGP_rate2}
\end{equation}
where the interpolation parameters $T_c$ is set to $T_c = 170$ MeV and $\Delta T = 0.1T_c$ \cite{Monnai:2019vup}.
We use the photon emission rate of QGP which parametrizes the result of leading-order pQCD calculation
in the strong coupling constant $g_s$ \cite{Arnold:2001ms},
and the thermal photon emission rate in hadronic matter 
given in Refs.~\cite{Holt:2015cda,Turbide:2003si,Heffernan:2014mla}.
Integrating Eq.~(\ref{eq_had_QGP_rate2}) over the whole volume element of fluid, one can obtain thermal photon radiation yield from hydrodynamic expansion. 
Here, for simplicity we continue hydrodynamic evolution until $T_f=116\ \mathrm{MeV}$ which is below the switching temperature 
$T_{\rm SW}=150\ \mathrm{MeV}$ \cite{Paquet:2015lta, Miyachi-Nonaka}. 

\subsection{Radiative Hadronization}
We give a brief explanation of radiative hadronization.  For details, refer to Ref.~\cite{RadReCo}.  
In the quark recombination picture, a meson formation by radiative hadronization is written as a 2-to-2 process, 
\begin{equation}
  q+\bar q \to M+\gamma. 
 \label{radiative_hadronization}
\end{equation}
We model this as a 2-step process, 
$q + \bar{q} \rightarrow M^* \rightarrow M + \gamma$, 
picking up a quark and anti-quark ("preformed") state $M^*$ with the original ReCo model \cite{Fries:2003kq,Fries:2003vb}, and then letting it decay into a meson $M$ and a photon $\gamma$.   
We call this the radiative ReCo model.
Notice that we do not consider this preformed state as any physical resonance 
but just as an intermediate state in radiative meson production.

The number of the photons emitted in the formation  of  mesons is 
given by the product of the number of preformed states ${dN_{M_*}}/{d^3P}$ and 
the photon distribution emitted from a preformed state $E_\gamma {dn_\gamma(k; M_*,P)}/{d^3k}$, 
\begin{equation}
E_\gamma \frac{d N_\gamma}{d^3k}=\kappa \int dM_*\, \varrho(M_*)\int d^3P 
\left(\frac{dN_{M_*}}{d^3P}\right)\left(E_\gamma \frac{dn_\gamma(k; M_*,P)}{d^3k }\right). 
\label{photon_distribution}
\end{equation}
Here, $\varrho(M_*)$ is an invariant mass  distribution of the preformed states. 
In this paper we use $\varrho(M_*)= \delta(M_*-(m_1+m_2))$  with  $m_1$ and $m_2$ being constituent quark masses. 
In Eq.~(\ref{photon_distribution}), the overall factor $\kappa$ is introduced for reflection of 
possible other effects on radiative hadronization. 
We will determine $\kappa$ by comparison with the experimental data. 

\section{Numerical Results}
\begin{figure}[tb]
\begin{minipage}{0.45\linewidth}
\centerline{%
\includegraphics[width=6cm]{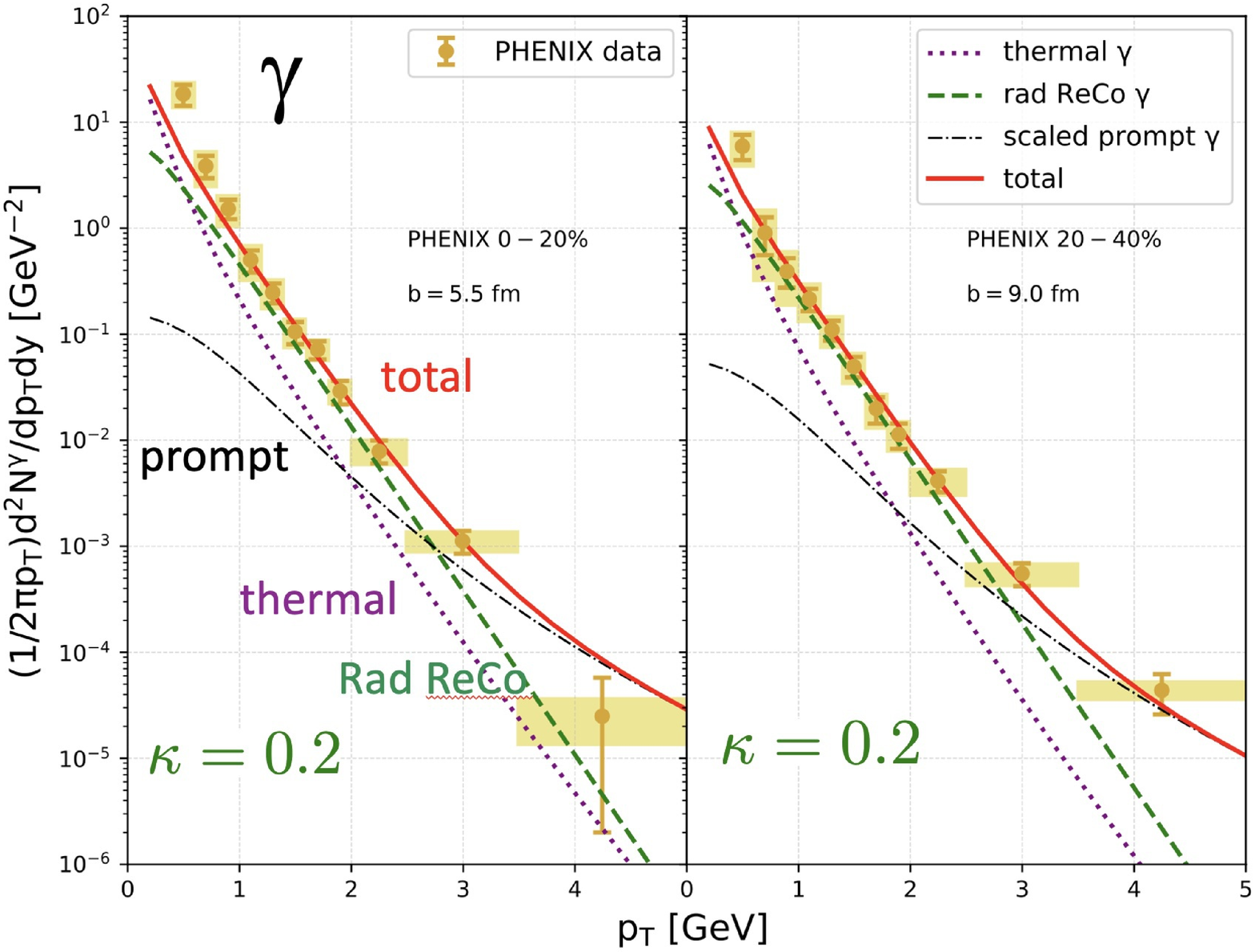}}
\caption{Transverse momentum distributions of direct photons 
for impact parameter $b=5.5$ fm (left) and 9.0 fm (right). 
}
\label{fig:photon_centrality_rhic}
\end{minipage}
\hspace{0.8cm}
\begin{minipage}{0.45\linewidth}
\vspace{0.45cm}
\centerline{%
\includegraphics[width=6.5cm]{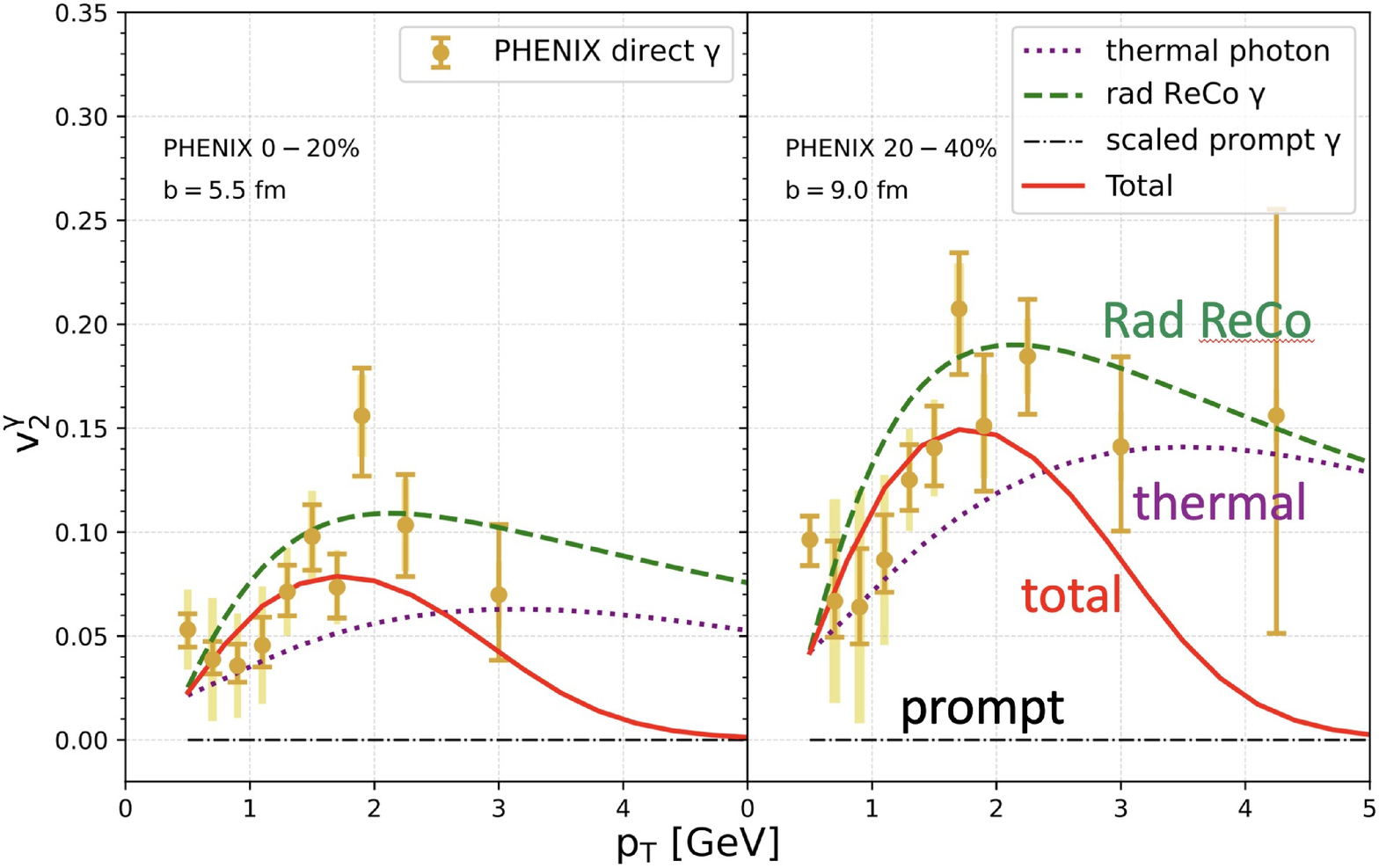}}
\caption{Elliptic flow coefficient $v_2^\gamma$ of the direct photons for impact parameter
$b=5.5$ fm (left) and 9.0 fm (right). 
}
\label{fig:v2_photon_rhic}
\end{minipage}
\end{figure}

In Fig.~\ref{fig:photon_centrality_rhic}
we show the $p_T$ spectra of the photons for $b=5.5$ fm (left panel) and $b=9.0$ fm (right panel) in 
$\sqrt{s_{NN}}=200$ GeV Au+Au collisions at RHIC. 
The red solid lines stand for total photons which consist of the thermal photons (purple dotted lines),  radiative pion production (green dashed lines) and the prompt photons (black dot-dashed lines).
Our estimate of the thermal photon contribution is smaller than the PHENIX data \cite{Adare:2014fwh, PHENIX:2022rsx}, which is  consistent with other hydrodynamic model studies \cite{Paquet:2015lta}. 
Regarding prompt photon production in AA collisions, 
we use the empirical fit of the photon distribution in pp collisions, $a_1 (1+p_T^2/a_2)^{a_3}$ ($a_{1,2,3}$ are constants),
as is done by PHENIX \cite{Adare:2014fwh}.
We set the normalization of the radiative ReCo model to $\kappa=0.2$ so that the sum of the three photon contributions 
reproduces the observed photon yield for $p_T < 3$ GeV.
We notice that the photon yield from the radiative ReCo model is estimated to be several times larger than that from the thermal radiation. 

In Fig.~\ref{fig:v2_photon_rhic} we show $v_2^\gamma(p_T)$ (red solid) of the total photon  as well as those of the thermal photon  (purple dotted) and of the radiative ReCo model (green dashed), separately.
In addition, we presumed the prompt photons have no collective flow (black dot-dashed).
The thermal photons have a nonzero $v_2^\gamma$ but its value is systematically below the observed values \cite{Adare:2014fwh}.  
On the other hand, the photons from the radiative ReCo model have $v_2^\gamma$ as large as the pion $v_2$, and its $p_T$ dependence is almost the same as that of the pions.
Since the photon yield of the radiative ReCo model is estimated to be several times larger than the thermal photon yield, the resultant $v_2^\gamma(p_T)$ of the total photons is close to that of the radiative ReCo model component and is consistent with the data for $p_T \lesssim  2$ GeV, albeit a large uncertainty. 
At larger $p_T$, the prompt photons dominate and the flow is suppressed.

\begin{figure}[htb]
\begin{minipage}{0.45\linewidth}
\centerline{%
\includegraphics[width=6.3cm]{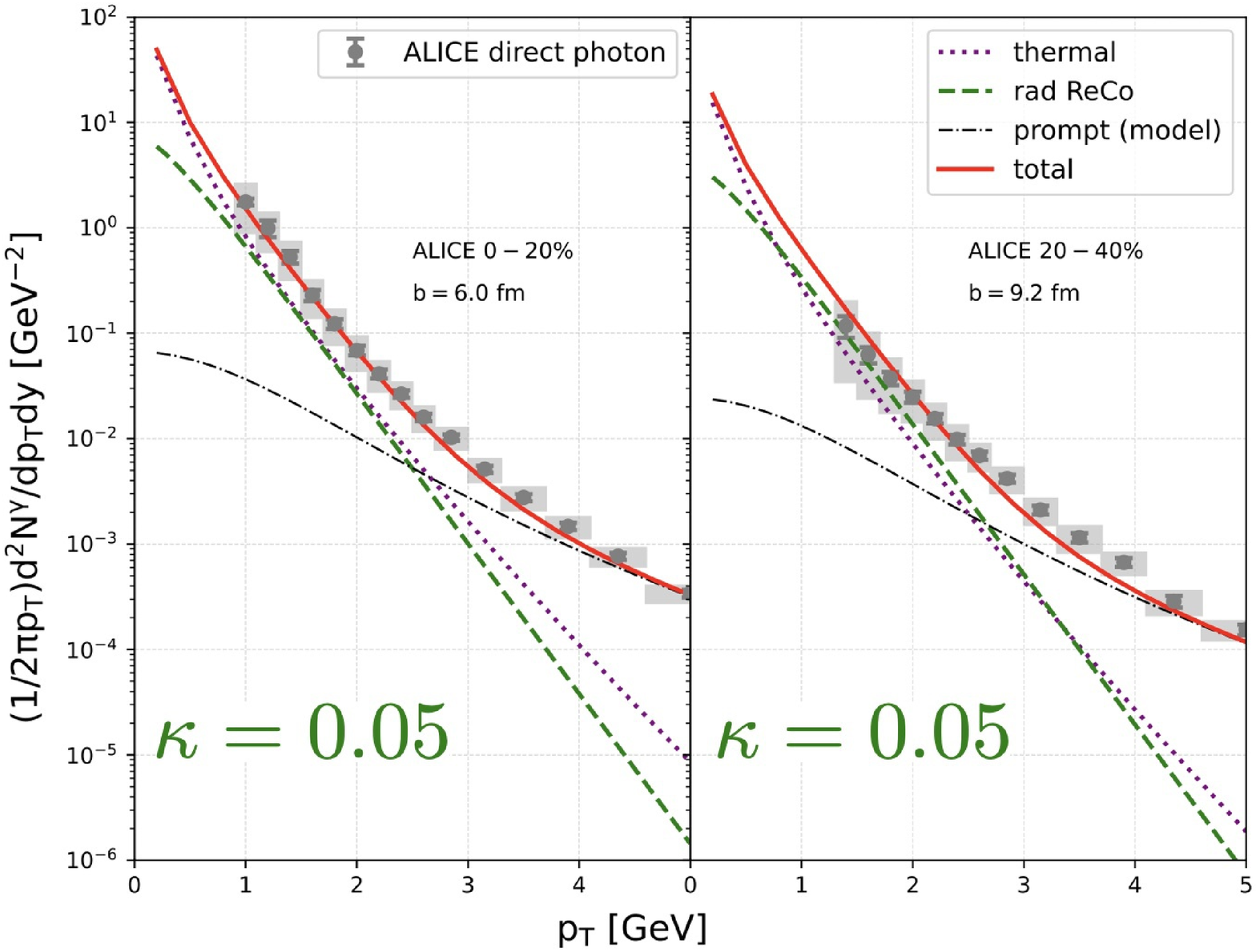}}
\caption{Transverse momentum distributions of direct photons 
for $b=6.0$ fm (left) and  $b=9.2$ fm (right). 
}
\label{fig:pt_lhc}
\end{minipage}
\hspace{0.8cm}
\begin{minipage}{0.45\linewidth}
\vspace{0.5cm}
\centerline{%
\includegraphics[width=6.5cm]{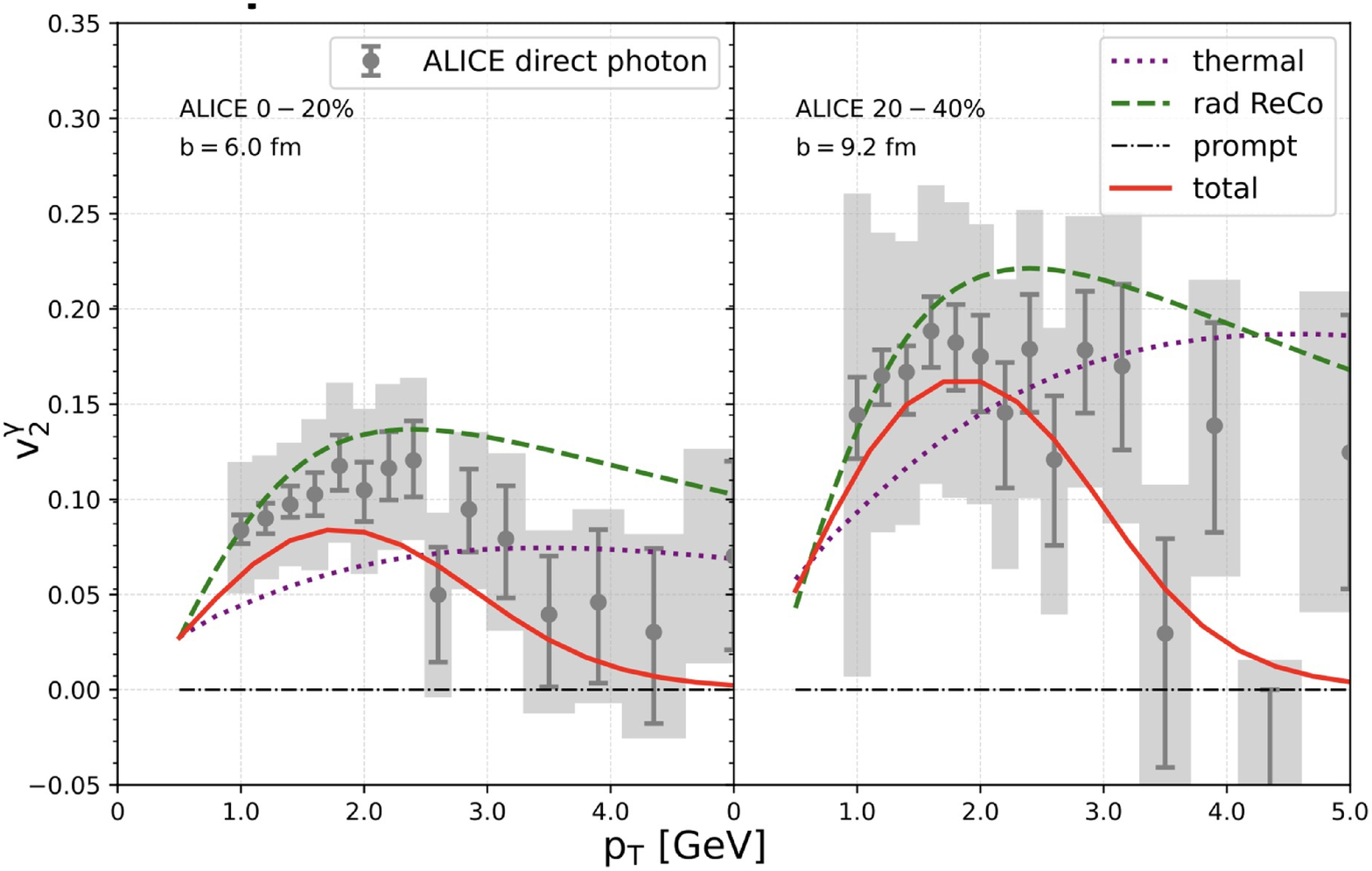}}
\caption{ 
Elliptic flow $v_2$ of direct photons (red solid) for $b=6.0$ fm (left) and 9.2 fm (right).
   } 
\label{fig:v2_lhc}
\end{minipage}
\end{figure}

Now we move on $\sqrt{s_{NN}}=2.76$ TeV Pb+Pb collisions at the LHC. 
In Fig.~\ref{fig:pt_lhc} 
we compare the photon $p_T$ distribution of our model at $b=6.0$ (9.2) fm
with the experimental data \cite{Abelev:2013vea} at 0--20 (20--40) \% centrality.
Unlike in the RHIC case, the thermal photon yield (purple dotted) is not far off the observed data in $1 < p_T < 2$ GeV,
and we can reasonably fit the data in the region $p_T \lesssim 2$ GeV
by adding the photons from the radiative ReCo model (green dashed) with the normalization factor $\kappa = 0.05$. 
We use the same model of photon distribution $a_1 (1+p_T^2/a_2)^{a_3}$ in pp collisions as before, 
setting the infrared cutoff $a_2=4$ GeV$^2$ and
tuning the parameter $a_1=1.2\times 10^{-2}$ GeV$^{-2}$ and $a_3 =-2.7$ to fit the available pp-collision data around $p_T \sim 10$ GeV at $\sqrt{s}=8$~TeV \cite{ALICE:2018mjj}.
We estimate the prompt photons in AA collisions by rescaling this model \cite{Adare:2014fwh}. 
Although we obtained a reasonable fit of the data in the low $p_T$ region with $\kappa=0.05$,  we don't have a clear explanation for the decrease of the $\kappa$ value
from the RHIC case.

Figure \ref{fig:v2_lhc} shows the $v_2^\gamma$ of photons for $b=6.0$ fm (left) and 9.2 fm (right).
Since in $1 < p_T < 2$ GeV the thermal radiation and radiative hadronization contribute almost equally to the photon yield,
the elliptic flow  $v_2^\gamma$ of the total photon yield becomes an average of the two sources and
lies just in between $v_2^\gamma$ of the thermal photon (purple dotted)  and $v_2^\gamma$ of radiative ReCo photons (green dashed) at lower $p_T$.
At higher $p_T$, it is dominated by prompt photon contribution, which we assume has the zero elliptic flow.

\vspace{-0.6cm}
\section{Summary and Discussions}
We have proposed ``radiative hadronization'' \cite{Fujii:2017nbv}  as an additional photon source and 
performed a comprehensive analysis for photon production at RHIC and the LHC. 
We embodied the radiative hadronization process by a two-step model, modifying  the original ReCo model \cite{Fries:2003kq,Fries:2003vb}. 
For numerical calculations, we considered three kinds of photon sources, prompt photons, thermal photons and photons from the radiative hadronization.

We have shown that the radiative ReCo model can account for a significant fraction of the total photon yield, 
while it comes from only a small fraction of total pion yields.
At the same time, we succeeded to fit the strong elliptic flow observed at RHIC, 
by adding the photons from the radiative ReCo model with the overall factor $\kappa=0.2$. 
For the LHC data, we obtained a reasonable fit of the $p_T$ distribution and the elliptic flow  of photons in $ 1 < p_T < 4$~GeV, but with the smaller value of $\kappa=0.05$.
The  reason for this decrease of $\kappa$ is the fact that the estimated thermal photon yield 
is not far off the observed direct photon yield compared to the RHIC case.
The origin of this change is unclear and open for future study.

\vspace{-0.3cm}
\section*{Acknowledgments}
This work was supported in part by JSPS KAKENHI Grant Numbers, JP16K05343 and JP21K03568 (HF), and JP20H00156, JP20H11581 and JP17K05438 (CN), and JP19K03836 (KI).


\vspace{-0.6cm}
\bibliographystyle{unsrt}
\bibliography{RadReCo}

\begin{thebibliography}{10}

\bibitem{Okamoto:2017rup}
Kazuhisa Okamoto and Chiho Nonaka.
\newblock {Temperature dependence of transport coefficients of QCD in
  high-energy heavy-ion collisions}.
\newblock {\em Phys. Rev. C}, 98(5):054906, 2018.

\bibitem{Miyachi-Nonaka}
Katsunori Miyachi and Chiho Nonaka.
\newblock in preparation.

\bibitem{Monnai:2019vup}
Akihiko Monnai.
\newblock {Prompt, pre-equilibrium, and thermal photons in relativistic nuclear
  collisions}.
\newblock {\em J. Phys. G}, 47(7):075105, 2020.

\bibitem{Arnold:2001ms}
Peter~Brockway Arnold, Guy~D. Moore, and Laurence~G. Yaffe.
\newblock {Photon emission from quark gluon plasma: Complete leading order
  results}.
\newblock {\em JHEP}, 12:009, 2001.

\bibitem{Holt:2015cda}
Nathan P.~M. Holt, Paul~M. Hohler, and Ralf Rapp.
\newblock {Thermal photon emission from the system}.
\newblock {\em Nucl. Phys.}, A945:1--20, 2016.

\bibitem{Turbide:2003si}
Simon Turbide, Ralf Rapp, and Charles Gale.
\newblock {Hadronic production of thermal photons}.
\newblock {\em Phys. Rev. C}, 69:014903, 2004.

\bibitem{Heffernan:2014mla}
Matthew Heffernan, Paul Hohler, and Ralf Rapp.
\newblock {Universal Parametrization of Thermal Photon Rates in Hadronic
  Matter}.
\newblock {\em Phys. Rev. C}, 91(2):027902, 2015.

\bibitem{Paquet:2015lta}
Jean-Fran\c{c}ois Paquet, Chun Shen, Gabriel~S. Denicol, Matthew Luzum, Bj\"orn
  Schenke, Sangyong Jeon, and Charles Gale.
\newblock {Production of photons in relativistic heavy-ion collisions}.
\newblock {\em Phys. Rev. C}, 93(4):044906, 2016.

\bibitem{RadReCo}
Hirotsugu Fujii, Kazunori Itakura, Katsunori Miyachi, and Chiho Nonaka.
\newblock {Radiative hadronization: Photon emission at hadronization from
  quark-gluon plasma}.
\newblock arXiv:2204.03116(nucl-th).

\bibitem{Fries:2003kq}
R.~J. Fries, Berndt Muller, C.~Nonaka, and S.~A. Bass.
\newblock {Hadron production in heavy ion collisions: Fragmentation and
  recombination from a dense parton phase}.
\newblock {\em Phys. Rev. C}, 68:044902, 2003.

\bibitem{Fries:2003vb}
R.~J. Fries, Berndt Muller, C.~Nonaka, and S.~A. Bass.
\newblock {Hadronization in heavy ion collisions: Recombination and
  fragmentation of partons}.
\newblock {\em Phys. Rev. Lett.}, 90:202303, 2003.

\bibitem{Adare:2014fwh}
A.~Adare et~al.
\newblock {Centrality dependence of low-momentum direct-photon production in
  Au$+$Au collisions at $\sqrt{s_{_{NN}}}=200$ GeV}.
\newblock {\em Phys. Rev.}, C91(6):064904, 2015.

\bibitem{PHENIX:2022rsx}
U.~A. Acharya et~al.
\newblock {Nonprompt direct-photon production in Au$+$Au collisions at
  $\sqrt{s_{_{NN}}}=200$ GeV}.
\newblock 3 2022.

\bibitem{Abelev:2013vea}
Betty Abelev et~al.
\newblock {Centrality dependence of $\pi$, K, p production in Pb-Pb collisions
  at $\sqrt{s_{NN}}$ = 2.76 TeV}.
\newblock {\em Phys. Rev.}, C88:044910, 2013.

\bibitem{ALICE:2018mjj}
Shreyasi Acharya et~al.
\newblock {Direct photon production at low transverse momentum in proton-proton
  collisions at $\mathbf{\sqrt{s}=2.76}$ and 8 TeV}.
\newblock {\em Phys. Rev. C}, 99(2):024912, 2019.

\bibitem{Fujii:2017nbv}
Hirotsugu Fujii, Kazunori Itakura, and Chiho Nonaka.
\newblock {Photon emission at hadronization}.
\newblock {\em Nucl. Phys. A}, 967:704--707, 2017.

\end{thebibliography}
\end{document}